\documentclass[11pt]{article}
\usepackage{graphicx}

\begin{document}

\begin{center}
{\textbf{ INTERACTING LINEAR POLYMERS ON THREE--DIMENSIONAL
SIERPINSKI FRACTALS}}
\end{center}

\vspace{11pt}

\begin{center}
 \textit{Jelena Mari\v ci\' c ${{}^1}$ and
 Sun\v cica Elezovi\' c--Had\v zi\' c
${{}^2}$}\\
 ${}^1$ Physics Department, University
of Hawaii at Manoa
\\
 ${}^2$ Faculty of Physics, University of Belgrade
\end{center}

\vspace{22pt}

\noindent {\bf Abstract}

\vspace{11pt}

Using self--avoiding walk model on three--dimensional Sierpinski
fractals (3d SF) we have studied critical properties of
self--interacting linear polymers in porous environment, via exact
real--space renormalization group (RG) method. We have found that
RG equations for 3d SF with base $b=4$ are much more complicated
than for the previously studied $b=2$ and $b=3$ 3d SFs. Numerical
analysis of these equations shows that for all considered cases
there are three fixed points, corresponding to the
high--temperature extended polymer state, collapse transition, and
the low--temperature state, which is compact or semi--compact,
depending on the value of the fractal base $b$. We discuss the
reasons for such different low--temperature behavior, as well as
the possibility of establishing the RG equations beyond $b=4$.

 \vspace{22pt}

\noindent {\bf Introduction}

At low temperatures and in poor solvents linear polymers are in
compact globule state, while at high temperatures and in good
solvents they have extended coil configuration [1].
Self--interacting self--avoiding walk (SISAW) on a lattice is a
good model for linear polymer in solvent. If $-\varepsilon<0$ is
the attractive energy per pair of nearest--neighbors monomers (not
directly connected) and $w=$exp$(\varepsilon/k_BT)$ the
corresponding Boltzman factor, then for $N$--step SISAW  and $N\gg
1$ the following behavior of the mean-squared end-to-end distance
$\langle R^2_N\rangle$ is expected:
\begin{equation}
\langle R^2_N\rangle\sim N^{2\nu}\, .
\end{equation}
There is a critical value of interaction $w=w_c$, such that for
any $w<w_c$ critical exponent $\nu=\nu_{\mathrm SAW}$, which
corresponds to extended coil phase, whereas for any $w>w_c$
exponent $\nu$ has smaller value, equal to $1/d$ for
$d$--dimensional homogeneous lattices. The value $w=w_c$
corresponds to the collapse transition, with $\nu=\nu_c$, between
the low-- and high--temperature $\nu$ values:
$1/d<\nu_c<\nu_{\mathrm SAW}$.

On non--homogeneous lattices collapse transition also occurs, with
the different values of the exponent $\nu$. Three--dimensional
Sierpinski fractals (3d SFs) are nonhomogeneous lattices where
exact real space renormalization group (RG) method  can be applied
for studying asymptotic properties of SISAW [2,3,4].

\vspace{11pt}

\noindent {\bf Renormalization group scheme}

Each member of the 3d SF family is labelled by $b=2,3,4,\cdots$
and can be constructed recursively starting with a tetrahedron of
base $b$, containing $b(b+1)(b+2)/6$ unit tetrahedrons -- so
called generator $G_1(b)$ (see fig.1).
\begin{center}
\includegraphics[width=3cm]{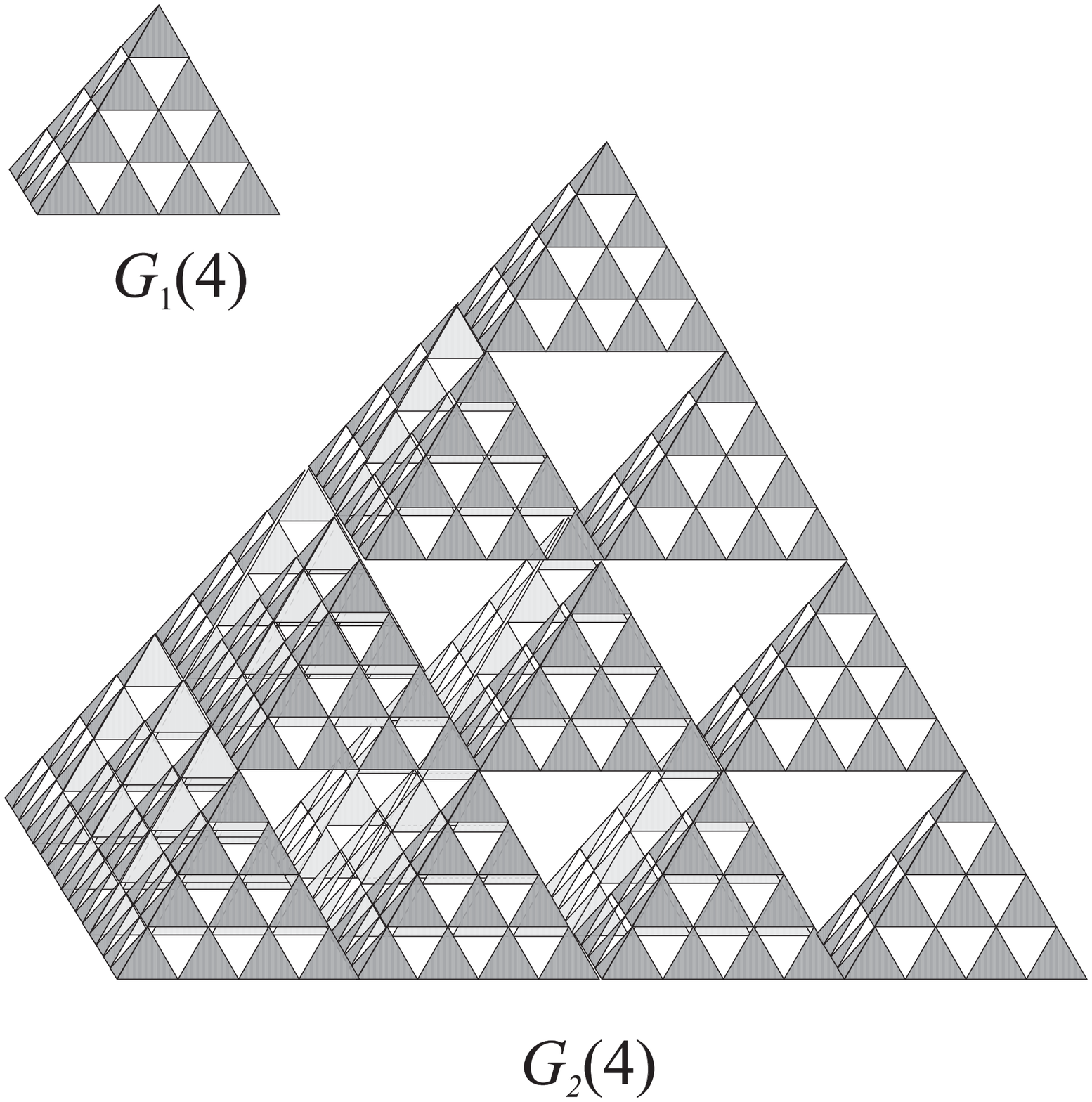}

Figure 1: First two stages of the construction of 3d SF with $b=4$
\end{center}
The subsequent fractal stages are constructed self--similarly, by
replacing each unit tetrahedron of the initial generator by a new
generator. To obtain the $l$th--stage fractal lattice $G_l(b)$,
which we shall call the $l$th order generator, this process of
construction has to be repeated $l-1$ times, and the complete
fractal is obtained in the limit $l\to\infty$. Fractal dimension
$d_f$ of 3d SF is equal to
\begin{equation}
d_f={{\ln [{{(b+2)(b+1)b}/ 6}}]/{\ln b}}\, .
\end{equation}

In this case RG parameters are restricted generating functions,
which are for the $G_l(b)$ equal to
\begin{equation}
 A^{(l)}(x,w)=\sum_{N,P}A_{N,P}x^Nw^P\, , \quad
B^{(l)}(x,w)=\sum_{N,P}B_{N,P}x^Nw^P\, ,
\end{equation}
where $x$ is weight of any step of a SISAW (fugacity), and
$A_{N,P}$ ($B_{N,P}$) is the number of $N$--step SISAWs with $P$
nearest--neighbor contacts, traversing the $G_l(b)$ once (twice)
-- see fig.2. For elementary tetrahedrons $G_0(b)$ one has
\begin{equation}
A^{(0)}(x,w)=x+2x^2w+2x^3w^3  \,,  \quad B^{(0)}(x,w)=x^2w^4\, .
\end{equation}
For any $b$ RG equations have the form
\begin{eqnarray}
A^{(l+1)}&=&\sum_{n,m}\mathcal A_{n,m}
\left(A^{(l)}\right)^n\left(B^{(l)}\right)^m\, , \nonumber \\
B^{(l+1)}&=&\sum_{n,m}\mathcal B_{n,m}
\left(A^{(l)}\right)^n\left(B^{(l)}\right)^m\, ,
\end{eqnarray}
where $\mathcal A_{n,m}$ ($\mathcal B_{n,m}$) is the number of
SISAW configurations consisting of $n$ branches of
"$A^{(l)}$"---type and $m$ branches of "$B^{(l)}$"---type,
traversing $G_{l+1}(b)$ once (twice).

\begin{center}
\includegraphics[width=4cm]{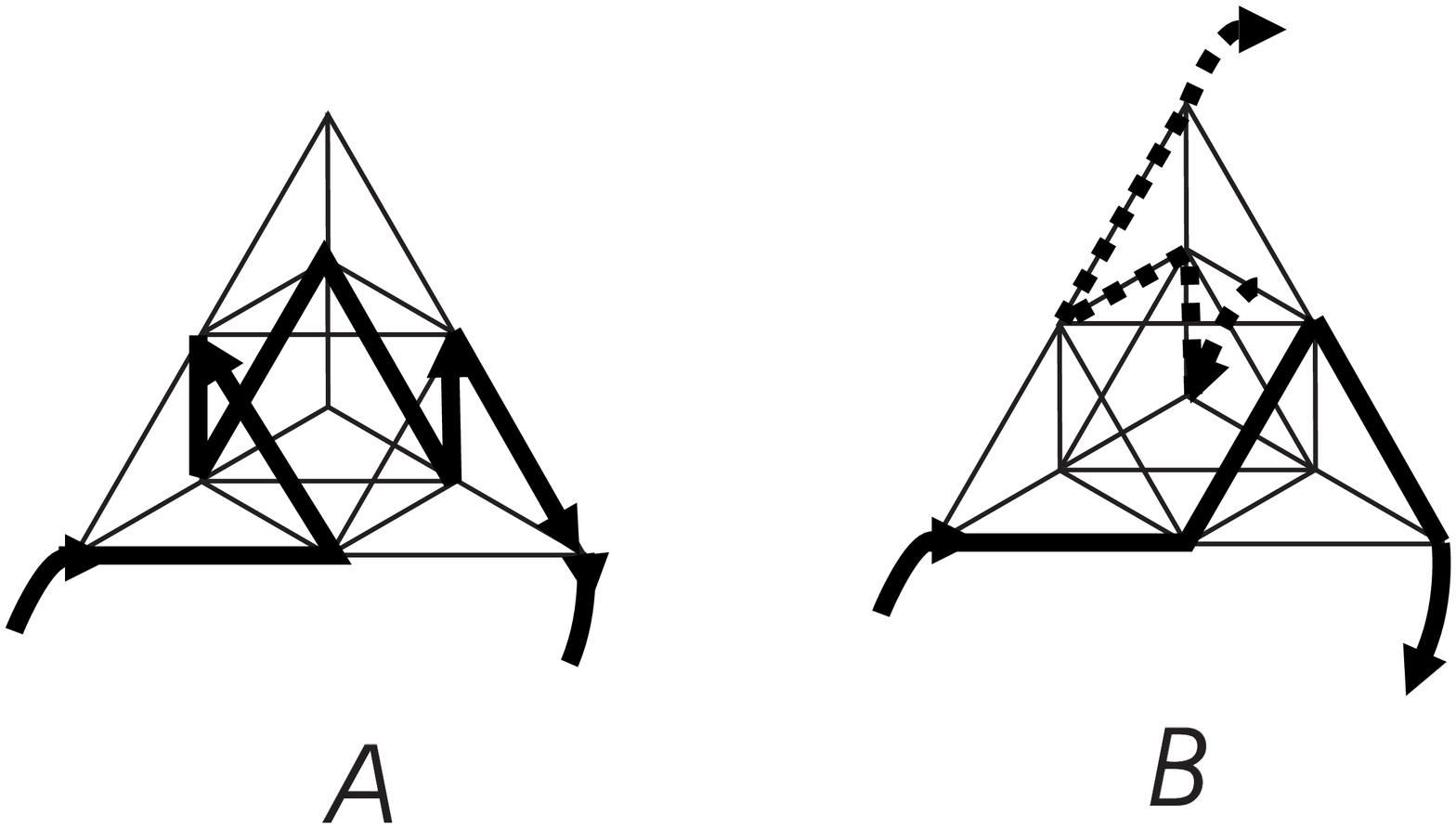}

Figure 2: Examples of $A$ and $B$ type of SISAW on $G_1(2)$
\end{center}

\noindent {\bf Analysis of the $b=4$ case}

As $b$ increases the number of possible SISAW configurations
quickly grows, which is the reason why we don't quote the
corresponding extremely cumbersome RG equations [4] for the
studied $b=4$ case here. Numerical analysis of these equations
shows that there are three physically relevant fixed points.

\noindent (1) The high--temperature fixed point ($A^*=0.289912,
B^*=0.0122393...$) is reached for any $w<w_c=5.028...$ and
corresponding critical fugacity $x=x^*(w)$. Linearizing RG
equations around this fixed point one gets only one relevant
eigenvalue $\lambda_1=8.6924...$, and the critical exponent $\nu$
is equal to $\nu_{\mathrm SAW}=\ln b/\ln\lambda_1=0.641076...$.

\noindent (2) The low--temperature fixed point ($A^*=0,
B^*=0.356883...=22^{-1/3}$) is reached when starting with $w>w_c$.
There is one relevant eigenvalue $\lambda_1=16$, and the critical
exponent is $\nu=1/2$.

\noindent (3) The tricritical fixed point ($A^*=0.19292...,
B^*=0.3388...$) corresponds to the collapse transition, when
$w=w_c$. In this case there are two relevant eigenvalues:
$\lambda_1=15.4194...$ and $\lambda_2=5.52427...$. The critical
exponent $\nu$ is equal to $\nu_c=\ln b/\ln\lambda_1=0.506756...$.

\vspace{11pt}

\noindent {\bf Discussion}

Values of $\nu$ for SISAW on all so far studied 3d SF [2-4] are
depicted on fig. 3. One can see that for any fractal $\nu$ has its
largest value in the high--temperature phase, and that it
decreases as $b$ grows, approaching the corresponding Euclidean
value $\nu_{\mathrm{SAW}}^E\approx 0.588$ [5]. Respective $\nu$
values are smaller at the collapse transition, also decreasing
with $b$, being close to the collapse Euclidean value
$\nu_c^E=0.5$ [1]. Low--temperature behavior is more interesting.
In contrast to the $b=2$ case for which $\nu=1/d_f$ (meaning that
polymer densely fills the underlying lattice, as on homogeneous
lattices), critical exponent $\nu$ for $b=3$ and $b=4$ is larger
than $1/d_f$, indicating that polymer is in {\em semi--compact},
rather than compact phase. This is due to the topological
frustration, which is not the same for $b=3$ and $b=4$, being also
the reason for strange behavior of the low-temperature RG fixed
point: $B^*$ is equal to $22^{-1/3}$ for both $b=2$ and $b=4$,
whereas for $b=3$ it is $B^*=\infty$ (of course, $A^*=0$ for any
$b$). It would be interesting to find out what happens beyond
$b=4$, but unfortunately the number of SISAW configurations grows
extremely fast with $b$. Estimated computer time required for
enumeration and sorting SISAWs for $b=5$ is not accessible, so one
should search other methods for calculating $\nu$.

\begin{center}
\includegraphics[height=5cm]{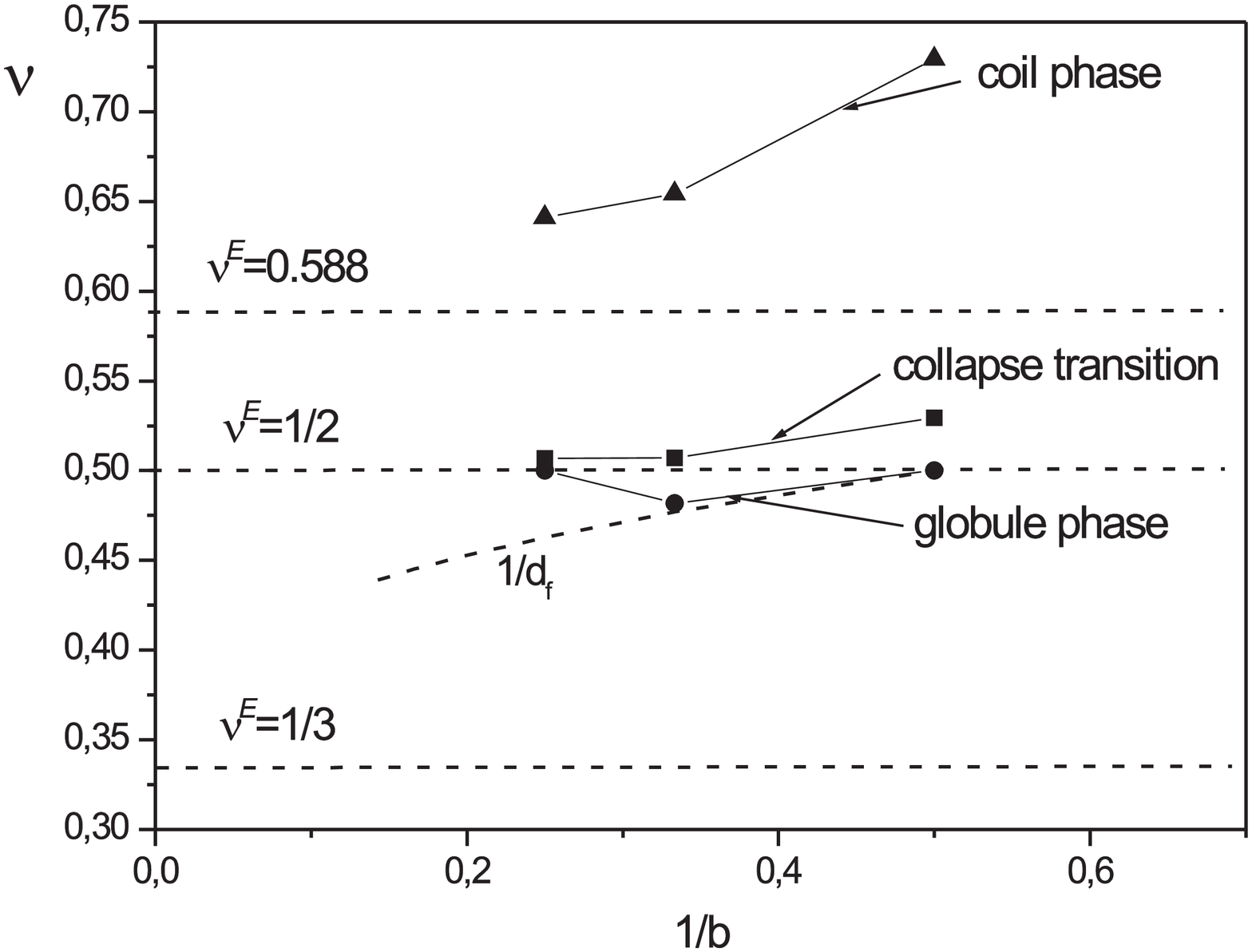}

Figure 3: Critical exponent $\nu$ for 3d SFs with $b=2,3$, and 4
\end{center}

\noindent {\bf References}

\noindent [1] P.G. de Gennes, "Scaling Concepts in Polymer
Physics" (Itacha, NY: Cornell University Press, 1979)

\noindent [2] D. Dhar and J. Vannimenus, J.Phys. A 20 (1987) 199

\noindent [3] M. Kne\v zevi\' c and J. Vannimenus, J.Phys. A 20
(1987) L969

\noindent [4] J. Mari\v ci\' c, "Random Walks on
Three--dimensional Sierpinski Fractals" -- diploma work (Faculty
of Physics, University of Belgrade, 1998)

\noindent [5] D. Lee et al., J.Stat.Phys. 80 (1995) 661; D.
MacDonald et al., J.Phys. A 33 (2000) 5973

\end{document}